\documentclass[reprint, amsmath, amssymb, aps]{revtex4-1}

\usepackage{color}
\usepackage{graphicx}
\usepackage{subfigure}
\usepackage{algorithm2e}
\usepackage{dsfont}
\usepackage{amsfonts}
\usepackage{amsthm}
\usepackage{booktabs}
\usepackage{extarrows}
\usepackage{multirow}

\def\>{\ensuremath{\rangle}}
\def\<{\ensuremath{\langle}}
\def\ra{\ensuremath{\rightarrow}}

\newcommand {\supp } {{\rm supp}}
\newcommand {\conf } {{\rm conf}}

\begin{document}

\title{Quantum Privacy-Preserving Data Mining}

\author{Shenggang Ying$^1$, Mingsheng Ying$^{1,2}$ and Yuan Feng$^1$}

\affiliation{$^1$Centre for Quantum Computation and Intelligent Systems, University of Technology Sydney, NSW 2007, Australia\\
$^2$State Key Laboratory of Intelligent Technology and Systems, Department of Computer Science and Technology,
Tsinghua University, Beijing 100084, China}

\date{\today}

\begin{abstract}
Data mining is a key technology in big data analytics and it can discover understandable knowledge (patterns) hidden in large data sets. Association rule is one of the most useful knowledge patterns, and a  large number of algorithms have been developed in the data mining literature to generate association rules corresponding to different problems and situations. Privacy becomes a vital issue when data mining is used to sensitive data sets like medical records, commercial data sets and national security.
In this Letter, we present a quantum protocol for mining association rules on vertically partitioned databases. The quantum protocol can significantly improve the privacy level preserved by known classical protocols and at the same time it can exponentially reduce the computational complexity and communication cost.
\end{abstract}

\maketitle

\textit{I. Introduction}.--- Numerous data mining techniques have been developed for discovering various knowledge patterns, using methods from statistics, machine learning, artificial intelligence and database systems. In this Letter, we focus on mining of association rules, one of the simplest but most useful knowledge patterns. An association rule is a relation between two disjoint sets of items. Suppose $I$ is a set of items. An association rule is a probabilistic implication $X\Rightarrow Y$, where $X$ and $Y$ are two disjoint subsets of $I$ with both the support (frequency) of $X\cup Y$ and the confidence of $X\Rightarrow Y$ being greater than preset values. The task of mining association rules is to find all such rules in a large database \cite{TanSK2006}.
Mining association rules or association analysis was first introduced in \cite{AgrawalIS1993} to find significant and useful relations between sets of items among a large collection of transactions in a supermarket. Since then, it has been very successfully applied in various areas including bioinformatics, medical diagnosis, market basket problem, inventory control, risk management, Web mining, telecommunication networks, earth science, scientific data analysis \cite{KotsiantisK2006,TanSK2006}.

A key issue in the real-world applications of these techniques is how to protect privacy in data mining. In the last 15 years, several privacy-preserving algorithms for mining association rules have been proposed \cite{SathiyapriyaS2013}. Two typical scenarios of privacy-preserving data mining are: (1) Discover statistical knowledge (e.g. frequency and confidence) with individual information being preserved; for example an algorithm was suggested in \cite{EvfimievskiSAG2004} for mining customers' behaviour in a supermarket while the personal behaviour of a customer is preserved. (2) Mine on a database that belongs to two or more parties, called a vertically partitioned database, but protect each party's privacy \cite{AgrawalES2003,VaidyaC2002,VaidyaC2005}.

In this Letter, we propose a quantum privacy-preserving method for mining association rules on vertically partitioned databases. Our method is based on the model of Quantum Random Access Memory (QRAM), a quantum generalization of Random Access Memory (RAM). QRAM was first introduced in \cite{GiovannettiLM2008qram}, and its physical architecture was proposed in \cite{GiovannettiLM2008Arch}. It was applied to define a quantum support vector machine for big data classification that can achieve an exponential speedup over classical algorithms \cite{RebentrostML2014}. Another interesting application of QRAM is quantum private queries \cite{GiovannettiLM2008PQ} that can detect whether the database provider cheats. The method presented in this Letter provides a new application of QRAM to one of the most researched problems in big data analytics, namely data mining association rules. This method can preserve the privacy much better than known classical protocols, and meanwhile exponentially reduce the computational complexity and communication cost.

\textit{II. Privacy-Preserving Mining of Association Rules}.--- Let us first formally define the problem of mining association rules in a database.
Suppose $I=\{I_1,\cdots,I_k\}$ is a set of items (or attributes), and $D=\{D_0,\cdots,D_{N-1}\}$ is a database, where each $D_i\subseteq I$ represents a transaction, which is, e.g. the set of items that one customer buys in one purchase. An itemset $X$ is a subset of $I$, and its support is defined as $\supp(X) = |\{j:X\subseteq D_j\}|/N$. The intuitive meaning of $\supp(X)$ is the frequency that $X$ appears in all transactions. A rule is a relation $X\Rightarrow Y$, meaning that itemset $X$ implies itemset $Y$, where $X$ and $Y$ are both itemsets with $X\cap Y=\emptyset$. Its confidence is defined as $\conf(X\Rightarrow Y) = \supp(X\cup Y)/\supp(X)$, which is actually the conditional probability $\Pr(Y|X)$. The mining problem is as follows: Given $s>0$ and $c>0$, find all pairs $(X,Y)$ with $X\cap Y = \emptyset$, $\supp(X\cup Y)>s$ and $\conf(X\Rightarrow Y)>c$. This problem can be solved in two steps: firstly, we find all itemsets $Z$ such that $\supp(Z)>s$, called frequent itemsets; secondly, we find all disjoint partitions $X\cup Y=Z$ with $\conf(X\Rightarrow Y)>c$, for every frequent itemset $Z$.
Both of the steps can be done by a level-wise algorithm; for example the \textit{Apriori} algorithm \cite{AgrawalS1994} first generates all 1-itemsets, i.e., itemsets with only 1 item. Then it generates all $m$-itemsets from all $(m-1)$-itemsets.

Now assume that the database $D$ is vertically partitioned into two disjoint parts $D^1$ and $D^2$ belonging to two parties: the set $I$ of items (or attributes) is divided to $I^1=\{I_1,\cdots,I_{l}\}$ and $I^2=\{I_{l+1},\cdots,I_{k}\}$ $(1\leq l<k)$. Alice owns (sub-)database $D^1=\{D^1_0,\cdots,D^1_{N-1}\}$ of $I^1$, and Bob owns (sub-)database $D^2=\{D^2_0,\cdots,D^2_{N-1}\}$ of $I^2$, where each transaction $D_j$ in $D$ is divided into $D^1_j\subseteq I^1$ and $D^2_j\subseteq I^2$ with $D_j=D^1_j\cup D^2_j$.
Then the privacy-preserving task is to mine association rules from $D^1$ and $D^2$ (e.g. compute the support of an itemset $Z\subseteq I$ with $Z^1=Z\cap I^1 \neq\emptyset$ and $Z^2=Z\cap I^2\neq\emptyset$), but Alice and Bob's private information must be preserved; that is, Bob (resp. Alice) is not allowed to know either the exact value of $D_j^1$ (resp. $D_j^2$) or whether $D_j^1\supseteq Z^1$ (resp. $D_j^2\supseteq Z^2$).

Two classes of protocols have been developed in the literature for privacy-preserving data mining on vertically partitioned databases. The first class adds noise to the data before transmitted, but the noise will be removed from the final results \cite{VaidyaC2002}. The second class employs hash functions and encryptions \cite{AgrawalES2003,VaidyaC2005}.

\textit{III. Quantum Database}.--- The aim of this Letter is to present a quantum protocol for privacy-preserving data mining. Before doing it, let us develop a quantum database model based on Quantum Random Access Memory (QRAM) \cite{GiovannettiLM2008qram}.
A QRAM can realize the following transform:
$$\sum \alpha_j|j\>|0\>\ra\sum \alpha_j|j\>|\tau_j\>,$$
where $\tau_j$ is the content in the $j$-th memory cell of the QRAM; that is, if we input the address qubits $|\psi\>=\sum \alpha_j|j\>$ with some blank data qubits, the QRAM outputs a superposition of some orthogonal quantum states which represent both address and data.

Note that in \cite{GiovannettiLM2008Arch}, the query of a QRAM is physically implemented by a set of controlled-NOT gates. As a consequence, querying the database twice with the same query will erase the data obtained in the first query. This feature will play a key role in our protocol.

Now we see how to store the partition $D^1$ and $D^2$ of database $D$ into QRAMs. We use a boolean string $d_j=d_{j,1}\cdots d_{j,k}$ to represent each transaction $D_j$, where $d_{j,i}=1$ if and only if $I_i\in D_j$ for every $1\leq i\leq k$. Furthermore, the string $d_j$ is divided into two substrings: $d_j^1=d_{j,1}\cdots d_{j,l}$ for Alice and $d_j^2 = d_{j,l+1}\cdots d_{j,k}$ for Bob. At Alice's (resp. Bob's) side, $d_j^1$ (resp. $d_j^2$) is stored in QRAM $\mathit{QD}^1$ (resp. $\mathit{QD}^2$). For convenience, we assume that $N=2^n$ for some integer $n$; otherwise we can add blank $d_j$.

\textit{IV. Quantum Protocol for Mining Association Rules}.---
We are now ready to present our quantum protocol. The basic idea is to employ the quantum counting algorithm \textbf{Count} in \cite{BrassardHT1998} to compute $\supp(Z)$ for any itemset $Z$. Recall that for a function $c:\{0,1\}^n\rightarrow\{0,1\}$, \textbf{Count} computes the number of $j$ with $c(j)=1$. The key component of \textbf{Count} is a controlled unitary called $G_C$:
$$|m\>\otimes|j\> \rightarrow |m\>\otimes G^m|j\>, \ 0\leq m\leq P-1$$ where $P$ is an integer that determines the precision of the computation, $G=-WU_0 W U$ is the Grover iteration, $W=H^{\otimes n}$ consists of $n$ Hadamard gates, and $U_0=\sigma_I^{\otimes n}-2|\vec{0}\>\<\vec{0}|$ with $\sigma_I$ being the identity operator and
$|\vec{0}\>=|0\>^{\otimes n}$, and
\begin{equation}\label{eq:U}
U: |j\>\rightarrow (-1)^{c(j)}|j\>
\end{equation}
is the oracle.


Note that to implement $G_C$, we need to perform $P$ controlled-$G$ operations, which in turn requires applications of controlled-$U$ operations. The main part of our protocol is devoted to the implementation of such a controlled oracle. For simplicity, we first implement $U$, the uncontrolled oracle, and then show how to adjust the protocol easily for the controlled one.



\textit{1. Implementing the oracle $U$}: The protocol starts with Alice. Suppose she holds the state $\sum\alpha_j|j\>$.

\textbf{Step 1}: Alice sends the state to Bob, and then Bob performs an encryption operator $E_B:$
$$\sum\nolimits\alpha_j|j\>\xrightarrow{E_B} \sum\nolimits\alpha_j|u(j)\>$$
on it, where $u$ is a bijection on $\{0,1\}^n$ chosen secretly by Bob. This choice of $E_B$ is rather arbitrary; it can be,
for instance, bit flip: $u(j)=j\oplus \lambda$ or addition: $u(j)=j+\lambda\mod N$ for a fixed $\lambda\in\{0,1\}^n$, or cyclic function: $u(j_1\cdots j_n) = j_2 \cdots j_n j_1$. These kinds of $E_B$ can be implemented by applying $O(n)$ $\sigma_{X}$, CNOT and Toffoli gates. The introduction of encryption operator is for the purpose of protecting Bob's privacy from Alice; we will discuss this issue later in more detail.

\textbf{Step 2}:  Bob queries his database $QD^2$ using the encrypted address qubits, applies $U_Z$ which is responsible for testing if the itemset $Z$ is contained in a transaction, and then query his database again to erase the data, thanks to the fact that QRAM is implemented by controlled-NOT gates. In effect, Bob makes the following transitions:
\begin{align*}
& \sum \alpha_j|u(j)\> \xrightarrow{Q_B}\sum \alpha_j|u(j)\> | d^2_{u(j)}\>\\
& \xrightarrow{U_Z} \sum \alpha_j|u(j)\> |d^2_{u(j)}\> |b_j\> \xrightarrow{Q_B}\sum \alpha_j|u(j)\> |b_j\>
\end{align*}
where $b_j=g_Z(d^2_{u(j)})$, $U_Z : |x\>|y\> \rightarrow |x\>|y\oplus g_Z(x)\>$, and $g_Z(x) = 1$ if and only if $X\supseteq Z^2$ with $X$ being the itemset represented by the boolean string $x$. For simplicity, we omit the auxiliary qubits in the above transitions when they are disentangled from the main systems and set back to their initial states.

Note that the query operation $Q_B$ costs $O(k\log N)$ time \cite{GiovannettiLM2008Arch}.
For $U_Z$, it can be implemented in $O(k)$ \footnote{To check whether $Z^2\subseteq D^2_j$, it suffices to check whether all the corresponding attributes in $D_j^2$ are 1, which can be reduced to check whether $x=2^m-1$ for $x\in\{0,1\}^m$ with $|Z^2|=m$. This can be done by quantum addition circuits. Firstly we add $1$ to $x$ to get $|x\>|x+1\>$. Secondly we check whether $x+1=2^m$ to get $|x\>|x+1\>|g\>$, where $g=1$ if and only if $x+1=2^m$. This step can be simply done by only one controlled-NOT gate, as $x+1\leq 2^m$. Finally we inverse the addition to disentangle the second system.}.
Thus the total time cost of this step is $O(k\log N)$.

\textbf{Step 3}: Bob sends back the address qubits and the qubit containing the $b_j$ information to Alice.
The task for Alice in this step is similar to Bob in Step 2:
\begin{align*}
& \sum \alpha_j|u(j)\> |b_j\> \xrightarrow{Q_A}\sum \alpha_j|u(j)\> |b_j\> | d^1_{u(j)}\>\\
& \xrightarrow{V_Z} \sum \alpha_j|u(j)\> |b_j\> |d^1_{u(j)}\> |a_j\> \xrightarrow{Q_A}\sum \alpha_j|u(j)\> |b_j\> |a_j\>
\end{align*}
where $a_j=f_Z(d^1_{u(j)})$ and $V_Z$ is defined similar to $U_Z$ but for $Z^1$.
Again, this step costs $O(k\log N)$ time.

\textbf{Step 4}: Alice computes $a_j b_j$ for each $j$ by first introducing an ancillary qubit initialised into $|-\>$, and then applying Toffoli gate with the ancillary qubit being target while the last two qubits from Step 3 being control qubits. In effect, she makes the transition:
\begin{align*}
  \sum \alpha_j|u(j)\> |b_j\> |a_j\>\xrightarrow{T_A}\sum\nolimits(-1)^{c_{u(j)}}\alpha_j|u(j)\> |b_j\> |a_j\>
\end{align*}
where $c_{u(j)}=a_j b_j$.

\textbf{Step 5}: Alice erases the information $a_j$ by performing the operations in Step 3 again. This leads to
\begin{align*}
  &\sum\nolimits(-1)^{c_{u(j)}}\alpha_j|u(j)\> |b_j\> |a_j\>\\
  &\hspace{5em} \xrightarrow{Q_A V_Z Q_A}  \sum\nolimits(-1)^{c_{u(j)}}\alpha_j|u(j)\> |b_j\>
\end{align*}
Actually, we can simplified the protocol a little bit by postponing the unquery of $Q_A$ in Step 3 to this step. This will save two applications of $Q_A$.

\textbf{Step 6}: Alice sends the systems obtained in the above step to Bob, who erases the information $b_j$ by performing the operations in Step 2 again:
\begin{align*}
  \sum\nolimits(-1)^{c_{u(j)}}\alpha_j|u(j) |b_j\>  \xrightarrow{Q_B U_Z Q_B}  \sum\nolimits(-1)^{c_{u(j)}}\alpha_j|u(j)\>
\end{align*}

\textbf{Step 7}: Bob undoes the encryption by performing $E_B^\dag$:
\begin{align*}
 \sum\nolimits(-1)^{c_{u(j)}}\alpha_j|u(j)\>  \xrightarrow{E_B^\dag}  \sum\nolimits(-1)^{c_{u(j)}}\alpha_j|j\>
\end{align*}
and sends the systems back to Alice. In this way the oracle $U$ defined in Eq.(\ref{eq:U}) has been implemented. Note that for the $E_B$ chosen in Step 1, $E_B^\dag$ can be implemented in time $O(n)$ as well.

We display the state evolution in the above steps in Table \ref{Tab:state}, which should help the reader to better understand the construction.

\begin{table}[b]
\caption{State evolution in Steps 1 through 7, where $U_A = Q_A V_Z Q_A$ and $U_B = Q_B U_Z Q_B$.  The subscriptions indicate which party the qubits belong to.\label{Tab:state}}
\begin{ruledtabular}
\begin{tabular}{l|l}
  Step  & State after the step\\\hline
  Init & $\sum_j\alpha_j|j\>_A$~~~~~~~~~~~~~~~~~~~~~~~~~~~~~~~~~~~\\
  1 $E_B$ & $\sum_j\alpha_j |u(j)\>_B$\\
  2 $U_B$ & $\sum_j \alpha_j|u(j),b_j\>_B$\\
  3 $U_A$ & $\sum_j \alpha_j|u(j),b_j,a_j\>_A$\\
  4 $T_A$ & $\sum_j (-1)^{c_{u(j)}}\alpha_j|u(j),b_j,a_j\>_A $\\
  5 $U_A$ & $\sum_j (-1)^{c_{u(j)}}\alpha_j|u(j),b_j\>_A $\\
  6 $U_B$ & $\sum_j (-1)^{c_{u(j)}}\alpha_j|u(j)\>_B $\\
  7 $E_B^\dag$ & $\sum_j (-1)^{c_{u(j)}}\alpha_j|j\>_A$\\
\end{tabular}
\end{ruledtabular}
\end{table}

\textit{2. Implementing the oracle controlled-$G$}. As said before, to make use of the quantum counting algorithm, we need to  implement the controlled-$G$ operation where $G=-WU_0 W U$. To this end, we need to modify the above protocol for $U$ to implement controlled-$U$. Fortunately, this turns out to be very easy. Denote by $q$ the control qubit of the controlled-$U$ we are going to implement. We only need to replace the Toffoli gate used in Step 4 by a controlled-Toffoli gate, with $q$ being the additional control qubit. The reason is, if Step 4 is missing, then the effects of the other 6 steps cancel each other.

Similarly, to implement controlled-$WU_0 W$, we need only to construct controlled-$U_0$. It is easy to see that the total complexity of implementing controlled-$G$ is still $O(k\log N)$.

\textit{3. Computing the support}: So far we described the protocol when it is initialised by Alice. The protocol initialised by Bob can be similarly defined, which implements controlled-$G'$ for Bob.
 Now the support of an itemset $Z$ can be computed as follows:
\begin{enumerate}
  \item Alice runs \textbf{Count} with controlled-$G$ as the oracle.
  \item Bob runs \textbf{Count} with controlled-$G'$ as the oracle.
  \item Alice and Bob announce their results $s_1$ and $s_2$. If $|s_1-s_2|<0.01s$, where $s$ is the preset threshold of support, they accept $\frac{s_1+s_2}{2}$ as the value of $\supp(Z)$. Otherwise they restart the protocol.
\end{enumerate}

One constraint for \textbf{Count} is that it can not distinguish between $\supp(Z)$ and $1-\supp(Z)$. Thus we assume $\supp(Z)<\frac{1}{2}$, as in \cite{BrassardHT1998}. This assumption is reasonable for most applications (e.g. mining customers' behaviour in a supermarket). By \cite[Theorem 5]{BrassardHT1998}, Bob can employ \textbf{Count} to compute $\supp(Z)$ with error smaller than $\frac{2\pi}{P}\sqrt{\supp(Z)}+\frac{\pi^2}{P^2}\leq \frac{2\pi}{P}+\frac{\pi^2}{P^2}$ and success probability greater than $\frac{8}{\pi^2}$. If we set $P\approx 2000/s$, the error will be smaller than $0.005s$. At the end of each round of this protocol, the probability that both $s_1$ and $s_2$ are in the interval $(s_0-0.005s,s_0+0.005s)$ are greater than $(\frac{8}{\pi^2})^2>0.64$, where $s_0=\supp(Z)$. Thus the terminating probability of each round is at least $0.64$, and the expected number of rounds is smaller than 2. Furthermore,
\begin{align*}
& \Pr(|\frac{s_1+s_2}{2}-s_0|>0.01s\ \wedge\ |s_1-s_2|<0.01s)
\\
& <(1-\frac{8}{\pi^2})^2<0.04.
\end{align*} Therefore, once the whole protocol terminates, it computes $\supp(Z)$ with error smaller than $0.01s$ and success probability greater than $\frac{0.64}{0.64+0.04}>0.9$.

\textit{4. Computing the confidence}: Since we can compute the supports $\supp(X)$ and $\supp(X\cup Y)$, the confidence $\conf(X\Rightarrow Y)=\supp(X\cup Y)/\supp(X)$ can be calculated with an error approximately equal to the sum of the errors of computing $\supp(X\cup Y)$ and $\supp(X)$.

\textit{V. Analysis of the Quantum Protocol}.--- We now show that the quantum protocol presented above significantly improves the privacy, complexity and communication cost of known classical algorithms.

\textit{1. Privacy}: We only consider the semi-honest model where both Alice and Bob follow the protocol, but may use the residual (classical as well as quantum) information after executing the protocol to extract the other party's private information \cite{AgrawalES2003,VaidyaC2005}.

Note that in \cite{AgrawalES2003,VaidyaC2005}, to compute the support of $Z$ secretly, Alice and Bob make use of commutative bijections $u_A$ and $u_B$, respectively, to encode the messages communicated between them: Alice first sends $u_A(S_1)=\{u_A(i):i\in S_1\}$ to Bob, where $S_1 = \{i:Z^1\subseteq D^1_i\}$. Then Bob sends back $u_B(u_A(S_1))$ to Alice.
Following the similar way but initialised by Bob, Alice and Bob can share $u_A(u_B(S_2))$ where $S_2 = \{i:Z^2\subseteq D^2_i\}$. Then they both compute
$$\supp(Z) = \frac{|S_1\cap S_2|}N = \frac{|u_B(u_A(S_1))\cap u_A(u_B(S_2))|}N.$$

A common choice of $u_A$ and $u_B$ is $u_A(x) \equiv x^{e_A} \mod p$ and $u_B(x)\equiv x^{e_B} \mod p$ where $p>N$ is a prime number, and $e_A$ and $e_B$ are coprime to $p-1$ and chosen from $\{3,5,\cdots,p-2\}$.  It has been proven secure if the attack is only allowed in \textit{polynomial time} with respect to $n$ and $k$  \cite{AgrawalES2003}. However, when \textit{exponential time} is permitted, then, say, Bob's data can be disclosed by Alice if she tries all possibilities of $e_B$, and compares the received $u_A(u_B(S_1))$ ($=u_B(u_A(S_1))$) with her own $u_A(S_1)$. For example if $p=11$, $e_A=9$, $e_B=3$ and $S_1=\{2,8\}$, then $u_A(S_1)= \{6,7\}$ and $u_A(u_B(S_1))=\{2,7\}$. Since only $w=3$ satisfies $\{6^w\mod 11,7^w\mod 11\}=\{2,7\}$, $e_B=3$ can be easily found, and Alice knows $S_2$ completely.

Compared to classical ones, our quantum protocol can better protect the participants' private information. Let us analyse the protocol initialised by Alice (the one we described in detail) as an example. From Table \ref{Tab:state}, Alice does not send her information $a_j$ to Bob, so Alice's privacy is protected perfectly. For Bob's privacy, note that at the end of the \textbf{Count} algorithm, Alice gets a (unnormalized) state
$$|\kappa_0\>=\sum\nolimits_{c_{u(j)}=0}\alpha_j|j\>\ \ \mbox{ or }\ \ |\kappa_1\>=\sum\nolimits_{c_{u(j)}=1}\alpha_j|j\>,$$
depending on the measurement result. Now the best way she can do to extract Bob's private information is to further measure the state according to the computational basis (that is, $\{|j\> : j=0, \ldots, N-1\}$). Let the measurement outcome be $j_0$. Then she knows $c_{u(j_0)}$ for sure. However, as $u$ is a bijection which is chosen secretly by Bob (Alice does not know, say, whether $u$ is a bit flip or an addition, how to choose $\lambda$ in each type, etc), this kind of information $c_{u(j_0)}$ is totally useless for Alice.
Note that if the encryption $u$ was not employed, then Alice gets the value $c_{j_0}$, and she can deduce whether $Z^2\subseteq D^2_{j_0}$ by using her own information of whether $Z^1\subseteq D^1_{j_0}$.

\textit{2. Time complexity}: The time cost of our protocol comes mainly from expected two ($\approx1/0.64$) rounds of $P$ calls of the controlled-$G$ operation,
which we have shown to be $O(k\log N)$-time implementable,
 and the Fourier transform in \textbf{Count}, which costs $O(\log^2 P)$ of time \cite{BrassardHT1998,NielsenC00}.
As we have set $P\approx 2000/s$ and the preset support threshold $s$ is usually a constant \cite{AgrawalIS1993},
 the total computational complexity of our protocol is $O(k\log N)$.

In contrast, a typical classical protocol needs $\Omega(N(k+poly(\log N)))$ time to compute the support of an itemset; for instance, the protocol given in \cite{AgrawalES2003} has the complexity $\Omega(kN+N\log^3 N)$. So, our quantum protocol is much faster than classical protocols.

\textit{3. Communication complexity}:
In our quantum protocol, communication only happens in Steps 1, 3, 6 and 7, and each time at most $n+1$ qubits are sent. Thus the communication cost of one application of controlled-$G$ is at most $4(n+1)=O(\log N)$. Furthermore, as the whole protocol calls controlled-$G$ $P$ times to compute $\supp(Z)$, the total communicate complexity is $O(\log N)$ qubits for each $Z$.

In contrast, in a classical protocol, since Alice and Bob have to share $u_B(u_A(S_1))$ and $u_A(u_B(S_2))$, the communication bits required is $\Omega(N\log N)$ \cite{AgrawalES2003,VaidyaC2005}.
A comparison of the time and communication complexities for the quantum protocol presented in this Letter and those of three popular classical protocols is given in Table \ref{Tab:comparison}.
\begin{table}[h]
\caption{Comparison of complexities for classical and quantum protocols. \label{Tab:comparison}}
\begin{ruledtabular}
\begin{tabular}{c|c|c}
  & Time &  Communication \\\hline
  Classical P. in \cite{AgrawalES2003} & \begin{tabular}{c}$\Omega(N\log^3 N+Nk)$\end{tabular} & $\Omega(N\log N)$ bits \\\hline
  Classical P. in \cite{VaidyaC2002} & \begin{tabular}{c}$\Omega(Nk)$\end{tabular} & $\Omega(N)$ bits \\\hline
  Classical P. in \cite{VaidyaC2005} & \begin{tabular}{c}$\Omega(poly(\log N)N$\\$+Nk)$\end{tabular} & $\Omega(N\log N)$ bits \\\hline
  \begin{tabular}{c}
  Quantum protocol
  \end{tabular} & $O(k\log N)$ & \begin{tabular}{c}
  $O(\log N)$  qubits
  \end{tabular}
\end{tabular}
\end{ruledtabular}
\end{table}

%

\textit{VI. Conclusion}.---In this Letter, we developed a quantum protocol for mining association rules on vertically partitioned databases. An analysis shows that the quantum protocol can better preserve the privacy than classical protocols, and at the same time it exponentially reduces the time and communication complexities. For the further research, we will explore the applications of quantum computation in data mining of other knowledge patterns. We will also consider how quantum computation can help in protecting privacy with other models of dishonesties.

\bibliography{qdm}

\end{document}